%% file: HIN-24-016_temp.tex
\pdfoutput=1
\documentclass[11pt,twoside,a4paper,cmspaper,final,collab]{cms-tdr}
\def\svnVersion{1b16fb8-D}\def\svnDate{2026/02/09}\def\cmsCernNoTag{CERN-EP-2025-260}\def\cmsCernDate{\today}\def\cmsMessage{Submitted to Physical Review Letters}
\begin{document}\cmsNoteHeader{HIN-24-016}

\cmsNoteHeader{HIN-24-016}
\title{Probing early parton emissions in heavy ion collisions using the Lund jet plane}

\newcommand{\pp}{\ensuremath{\Pp\Pp}\xspace}
\newcommand{\PbPb}{\ensuremath{\text{PbPb}}\xspace}
\newcommand{\delc}{\ensuremath{\Delta_\mathrm{c}}\xspace}
\newcommand{\jetpt}{\ensuremath{\pt^{\text{jet}}}\xspace}
\newcommand{\tform}{\ensuremath{t_\text{form}}\xspace}
\newcommand{\Lmed}{\ensuremath{L_\text{med}}\xspace}
\newcommand{\tz}{\ensuremath{t_0}\xspace}
\newcommand{\zg}{\ensuremath{z_{\mathrm{g}}}\xspace}
\newcommand{\Rg}{\ensuremath{R_{\mathrm{g}}}\xspace}
\newcommand{\lnD}{\ensuremath{\ln(1/\Delta)}\xspace}
\newcommand{\jsub}[1]{\ensuremath{j_\mathrm{#1}}\xspace}
\newcommand{\ja}{\ensuremath{\jsub{a}}\xspace}
\newcommand{\jb}{\ensuremath{\jsub{b}}\xspace}
\newcommand{\jab}{\ensuremath{\jsub{a,b}}\xspace}
\newcommand{\tpiT}{\ensuremath{2/(\pi T)}\xspace}
\newcommand{\Njets}{\ensuremath{N_{\text{jets}}}\xspace}
\newcommand{\mZ}{\ensuremath{m_\PZ}\xspace}
\newcommand{\JEWEL}{\textsc{jewel}\xspace}
\newcommand{\HYBRID}{\textsc{hybrid}\xspace}
\newcommand{\JETMED}{\textsc{hybrid}\xspace}
\newcommand{\JETSCAPE}{\textsc{jetscape}\xspace}

\renewcommand{\appendixname}{End Matter}

\date{\today}

\abstract{
In scattering experiments, high-virtuality partons, \ie, quarks and gluons, initiate a series of additional parton emissions to create collimated sprays of particles known as jets. This paper presents a measurement of the Lund jet plane (LJP) of high-energy jets produced in lead-lead (\PbPb) collisions and compares the results to data for proton-proton (\pp) collisions. The LJP is formed by iteratively declustering the constituents of a jet into consecutive emissions and recording the relative transverse momentum (\kt) and angle of the resulting emission with respect to its emitter. The angular distributions of two different \kt slices of the LJP are investigated for jets with radius parameter of 0.4 and transverse momentum in the range 200--1000\GeV. The \PbPb (\pp) data were recorded by the CMS experiment in 2018 (2017) and correspond to an integrated luminosity of 1.7\nbinv (301\pbinv) at a nucleon-nucleon center-of-mass energy of 5.02\TeV. The measurement was designed to test whether the earliest jet emissions are produced before the formation of the quark-gluon plasma (QGP) in \PbPb collisions. Within the experimental uncertainties, no significant difference is observed between the angular distribution of high-\kt emissions in \pp and \PbPb collisions, which is consistent with these emissions occurring early in the jet evolution, before substantial interaction with the QGP.
}

\hypersetup{
pdfauthor={CMS Collaboration},
pdftitle={Probing early parton emissions in heavy ion collisions using the Lund jet plane},
pdfsubject={CMS},
pdfkeywords={CMS, measurement, jet substructure, quarks, gluons, jets, Lund jet plane}}

\maketitle 

The quark-gluon plasma (QGP) is a high-temperature, strongly interacting phase of nuclear matter that is produced in ultrarelativistic heavy ion collisions~\cite{Busza:2018rrf,Adams:2012th}.
A fundamental goal of studying such collisions is to understand the macroscopic and microscopic characteristics of the QGP~\cite{Hayrapetyan:2897965}.
In these collisions, a hard-scattered quark or gluon (also known as a parton) can initiate a cascade of additional parton emissions that ultimately produces a collimated collection of particles known as a jet. Jets are produced in the early stages of the collision, and their fragmentation is intertwined with the time evolution of the QGP.
As a result of the jet-QGP interactions, which happen over a wide range of scales (from the high virtuality of the initial parton starting from hundreds of \GeV down to the QGP medium temperature of hundreds of \MeVns), jets experience energy loss (quenching) and alterations in their radiation pattern \cite{Cunqueiro:2021wls, Apolinario:2022vzg}.  

The dominant interaction mechanism between the partons and the QGP at short distance scales is expected to be gluon radiation induced by the medium~\cite{Gyulassy:1990ye,Baier:1996kr,Baier:1996sk}.
The QGP medium is characterized by a ``coherence angle'', \delc, or alternatively ``resolution length'', $L$~\cite{Casalderrey-Solana:2011ule,Mehtar-Tani:2010ebp}, which sets the minimum angular separation between two partons or subjets that can be resolved by the medium as independent color charges~\cite{Casalderrey-Solana:2012evi}. 
As the jet fragmentation evolves, the jet pushes QGP constituents in the jet direction, giving rise to a correlated background referred to as the medium response or the wake \cite{Cao:2022odi}.   

Jet substructure observables are used to study and characterize the aforementioned phenomena, with the aim of mapping them to the relevant energy and angular scales. 
This will allow a consistent picture of medium-modified jet evolution to be constructed, from which the microscopic properties of the QGP can be inferred. 
An underlying assumption in the theoretical description of a jet's evolution  in a heavy ion collision is that the process can be factorized into an early vacuum-like parton showering stage and a medium-modification stage. 
This assumption is based on timescale arguments, \ie, that a vacuum-like shower at higher virtualities is initialized before the formation of the QGP. 
Exact or partial factorization between vacuum and medium-modified processes is implemented in several jet quenching Monte Carlo (MC) event generators~\cite{Caucal:2018dla,JETSCAPE:2023hqn}. 
Other approaches implement interleaved medium and vacuum evolution~\cite{Zapp:2012ak}. 
The present analysis aims to experimentally test whether evidence of early vacuum-like processes can be observed in heavy ion collisions.

\begin{figure*}
    \centering
    \includegraphics[width=0.9\linewidth]{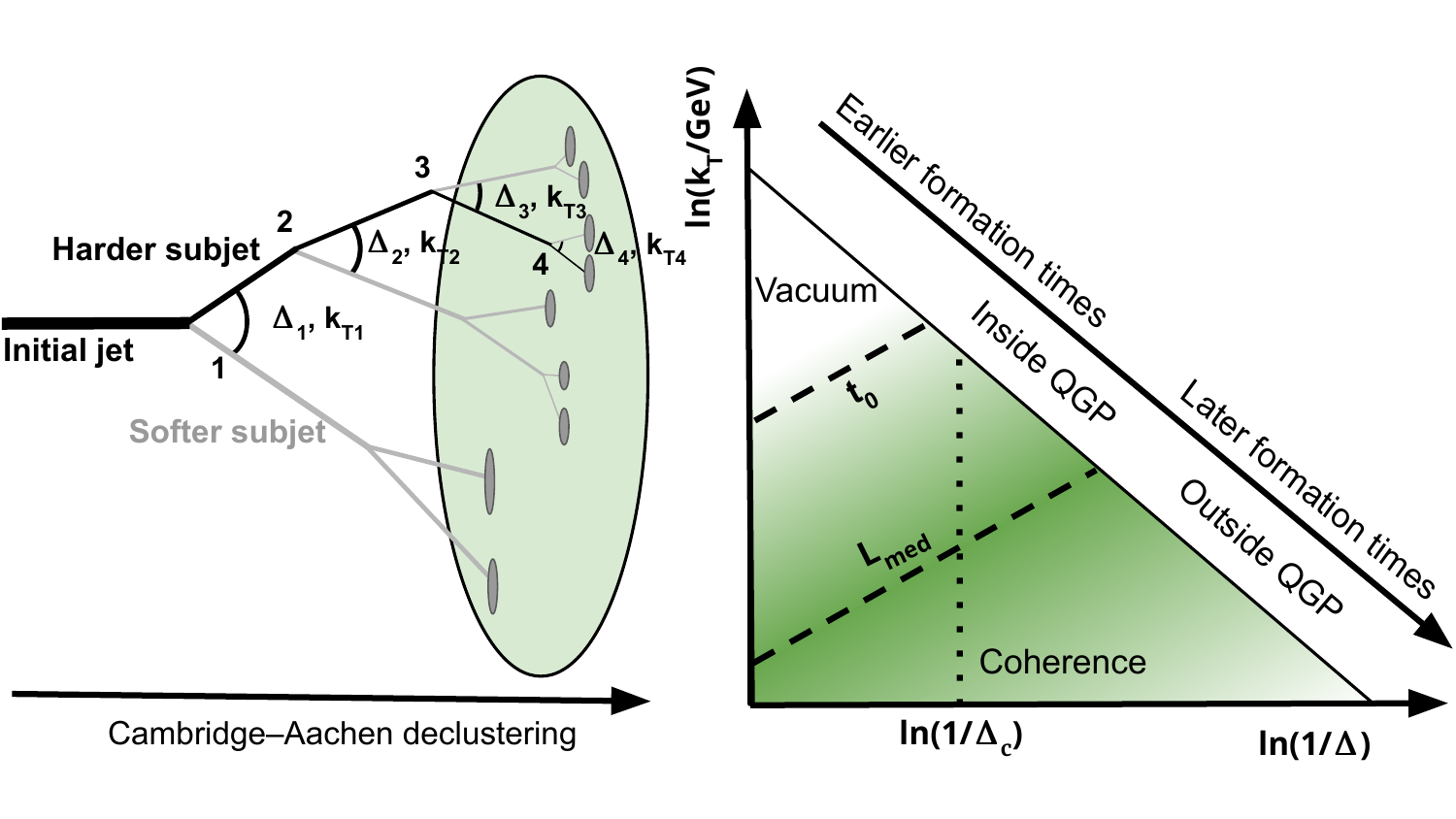}
    \caption{Left: Schematic diagram of the Cambridge--Aachen primary declusterings. The kinematics of the softer (grey) subjets in subsequent numbered emissions at each step from the harder subjet (bold black) is mapped onto the PLJP.
    Right: Distinct regions of the PLJP with the expected sensitivities of the jet shower evolution as the jet shower traverses the QGP, with maximum path length inside the medium \Lmed.
    Emissions with formation times $\tform > \Lmed$ are formed outside of the QGP, while emissions with $\tform < \tz$, where \tz is the hydrodynamic initialization time, are vacuum-like. 
    The typical values of these timescales are $\tz\sim0.6\unit{fm}$ and $\Lmed\sim 6\unit{fm}$.}
    \label{fig:lp_cartoon}
\end{figure*}

The primary Lund jet plane (PLJP)~\cite{Lifson:2020gua,Dreyer:2018nbf} provides a way of studying the jet shower.
The PLJP is a two-dimensional representation of the intrajet processes in which a parton emits two new partons which further fragment to form subjets. The properties of these emissions is experimentally accessible using the iterative Cambridge--Aachen (CA) reclustering algorithm \cite{Dokshitzer:1997in}. 
The constituents of a jet are reclustered using the CA algorithm, following the prescription in Ref.~\cite{Dreyer:2018nbf}.
Then, the CA pairwise clustering history is followed in reverse, as shown schematically in the left panel of Fig.~\ref{fig:lp_cartoon}.
Starting with the reclustered jet ${j}$, the last step of the CA clustering is undone, so that the original jet is declustered into the two subjets of the previous clustering step $j \to \ja+\jb$, where \ja is the harder subjet (the bold black line in Fig.~\ref{fig:lp_cartoon}, left panel), and \jb is the softer subjet (the grey lines stemming from the black line in Fig.~\ref{fig:lp_cartoon}, left panel) in the branching.
The softer subjet is treated as an emission in the PLJP and its angle, $\Delta$, and relative transverse momentum, \kt, with respect to the harder subjet, are determined in the Lund plane (Fig.~\ref{fig:lp_cartoon}, right panel). The observables are defined as
\begin{equation}
  \Delta = \sqrt{{(y^{\ja}-y^{\jb})^2+(\phi^{\ja}-\phi^{\jb})^2}},\quad \kt = \pt^{\jb} \Delta,
\end{equation}
where $y^{\jab}$ and $\phi^{\jab}$ represent the rapidity and the azimuthal angle of the subjets, respectively, and $\pt^{\jb}$ is the transverse momentum (\pt) of the emission relative to the beam axis.
Subjet \ja is then declustered in the next step. 
The formation time of emissions is approximated by the expression $\tform=2/(\kt\Delta)$~\cite{Cunqueiro:2023vxl}, leading to the earliest formed emissions occupying the upper left corner of the PLJP shown on the right side of Fig.~\ref{fig:lp_cartoon}.
The emission with the earliest formation time is not necessarily the first in the CA tree, as the CA algorithm is independent of the emission \pt.
The QGP generates two new scales as compared to quantum chromodynamics (QCD) in vacuum: the propagation length inside the QGP, and the angle \delc.
These two scales govern how the QGP-induced signal populates the Lund plane.

The PLJP has been explored in previous measurements at the LHC with proton-proton (\pp) collisions \cite{CMS:2023lpp,ATLAS:2020bbn,LHCb:2025mcq}. In lead-lead (\PbPb) collisions, the measurement of the Lund plane is hampered by the large underlying event (UE) consisting of particles not stemming from the hard interaction, \eg, particles produced in other nucleon-nucleon interactions in the collision.
This effect can be mitigated by selecting emissions from the PLJP using grooming algorithms such as soft-drop~\cite{Larkoski:2014wba} and dynamical grooming~\cite{Mehtar-Tani:2019rrk}. 
The distributions of the groomed opening angle \Rg, the momentum balance \zg, the mass and the scale \kt of the selected splittings, as well as the \kt distance have been measured \cite{Sirunyan:2017bsd,CMS:2018fof,ALICE:2019ykw,ALargeIonColliderExperiment:2021mqf,ATLAS:2022vii, ALICE:2024fip, ATLAS:2023hso}.

In this analysis, the proposal presented in Ref.~\cite{Cunqueiro:2023vxl} is followed, in which the angular structure of the PLJP formed by only the highest \kt emission in each jet is analyzed in separate ranges of \kt.
Selecting directly on \kt allows the separation of soft and hard energy scales. 
This is not the case for other measurements like \Rg and \zg, which span a wider range of \kt values (from a few hundreds of MeV up to tens of \GeV). 
The selection of very hard splittings is intended to access a kinematic region where the splittings within the jet shower, which is subsequently modified by the medium, are vacuum-like.

The analysis is based on \PbPb and \pp data collected by the CMS experiment in 2018 and 2017, at a nucleon-nucleon center-of-mass energy of 5.02\TeV, with integrated luminosities of 1.70\nbinv and 301\pbinv, respectively~\cite{CMS-PAS-LUM-20-002,CMS-PAS-LUM-19-001}.
Tabulated results are provided in the HEPData record for this
analysis~\cite{HEPData}.

The CMS apparatus~\cite{CMS:2008xjf,CMS:2309.05466} is a multipurpose, nearly hermetic detector, designed to trigger on~\cite{CMS:2020cmk,CMS:2016ngn} and identify electrons, muons, photons, and charged and neutral hadrons~\cite{CMS:2020uim,CMS:2018rym,CMS:2014pgm}. 
A global ``particle-flow'' (PF) algorithm~\cite{CMS:2017yfk} aims to reconstruct all individual particles in an event, combining information provided by an all-silicon inner tracker, a crystal electromagnetic calorimeter (ECAL), and a brass-scintillator hadron calorimeter (HCAL), operating inside a 3.8\unit{T} superconducting solenoid, with data from gas-ionization muon detectors embedded in the flux-return yoke outside the solenoid.
The HCAL (ECAL) provides pseudorapidity coverage up to $\abs{\eta}=3$ and is divided into a barrel and endcaps regions, covering $\abs{\eta}<1.3 (1.5)$ and $1.3 (1.5)<\abs{\eta}<3$, respectively.
Moreover, the hadron forward (HF) calorimeter extends the calorimeter range up to $\abs{\eta}<5.2$.

Events of interest are selected using a two-tiered trigger system. The first level, composed of custom hardware processors, uses information from the calorimeters and muon detectors to select events at a rate of around 30 (100)\unit{kHz} for \PbPb (pp) collisions~\cite{CMS:2020cmk}. 
The second level, known as the high-level trigger~\cite{CMS:2016ngn}, consists of a farm of processors running a version of the full event reconstruction software optimized for fast processing, and reduces the event rate to a few kHz before data storage. 

The collision centrality in heavy ion collisions is expressed as a percentile of the total inelastic hadronic cross section, following the Glauber methodology~\cite{Loizides:2017ack}, and is determined based on the transverse energy deposited in both HF calorimeters~\cite{CMS:2011iwn}. 
The events with close to 0\% centrality, known as ``central'' events, have the largest overlap of the two colliding nuclei, whereas glancing collisions between the two nuclei are referred to as ``peripheral'' events.
Hadronic collisions are selected by requiring at least two towers with an energy larger than 4\GeV in each of the HF calorimeters. 
In order to focus on medium modification effects, the analysis is restricted to events with centrality less than 30\%.
The events are also required to have a primary vertex reconstructed within 25\unit{cm} of the nominal interaction point along the beam direction, and within 0.15\unit{cm} in the transverse plane.

Jets are reconstructed by clustering PF candidates using the anti-\kt algorithm \cite{Cacciari:2008gp, Cacciari:2011ma} with a radius parameter $R=0.4$ and four-momentum equal to the vector sum of all particle momenta inside the jet.
The jets are required to fulfill the selection criteria of $185<\jetpt<1000\GeV$ and $\abs{\eta^{\text{jet}}}<2$.
The value of the lower \jetpt bound is chosen such that the trigger used to collect the high-\pt jets, requiring at least one calorimeter jet with energy above 100\GeV, is fully efficient.
The $\eta$ selection allows for the whole jet to be contained within the tracker acceptance.
In \PbPb collisions, the jet constituents are corrected for the UE contribution using the constituent subtraction method~\cite{Berta:2014eza}.
This approach removes or corrects jet constituents based on the average UE energy density, which is estimated using the procedure described in Ref.~\cite{Kodolova:2007hd}.
Because of fluctuations of the UE in the $\eta$--$\phi$ plane, residual background fluctuations smear the jet \pt and its internal structure \cite{ALICE:2012nbx}, which are corrected with unfolding.

In order to account for the detector response, jet energy scale (JES) corrections are applied  using independent \PbPb and \pp simulations, along with additional corrections for the imperfect modeling of the detector response \cite{CMS:2011shu}.
In addition, the jet energy resolution (JER) has been found to be smaller in data compared with simulation.
Corrections to the simulation to account for the JER differences are derived from dijet balancing studies performed in \pp collisions at $\sqrt{s} = 13\TeV$ in 2017--2018~\cite{CMS:2016lmd}.
To incorporate this effect, a Gaussian smearing is applied to the reconstructed by the detector (detector-level) \jetpt values in simulation to match the resolution in data.
As the jet emission of interest is the one with the highest \kt, it is also expected to carry a significant fraction of the jet momentum, since the maximum value of \kt is given by $\kt^{\text{max}} =  \jetpt\Delta/2$ (from kinematic arguments). 
Consequently, JES and JER corrections are applied both to the \jetpt and to the \kt of the hardest emission. 

Events generated by \PYTHIA{ 8.230} \cite{Sjostrand:2014zea} with the UE tune CP5~\cite{CMS:2019csb} and \HERWIG{ 6.521}~\cite{Corcella:478828, Corcella:586836} with the CH3 UE tune~\cite{Sirunyan:2020pqv} are used to calculate corrections to simulations and for making comparisons to the \pp data. 
For the \PbPb simulation, \PYTHIA events are embedded into a heavy ion UE produced with \HYDJET{ v1.9} \cite{Lokhtin:2005px}, to include the multiple nucleon-nucleon interactions occurring in the same \PbPb collision. This simulation is tuned to reproduce the global event properties in \PbPb collisions.
All generated samples are passed through a detailed simulation of the CMS detector using \GEANTfour~\cite{Agostinelli:2002hh}.

In heavy ion collisions, the UE alters the CA jet clustering pattern by smearing the momentum and direction of the subjets and by creating fake emissions. 
The contamination of these false emissions is stronger for low-\jetpt, and large-$R$ jets, and depends on the grooming algorithm~\cite{Mulligan:2020tim}.
To suppress this background, a cutoff of 3\GeV on the detector-level value of the \kt observable is applied. 

To facilitate comparisons with theoretical predictions and with other experiments, the detector-level distributions of $\Delta$, \kt and \jetpt are unfolded to the level of stable particles to remove detector- and UE-induced shifts.
These shifts are quantified using a six-dimensional response matrix constructed from the particle- and detector-level distributions of the observables, which share the same binning, with the addition of underflow and overflow bins at particle-level to allow for migrations coming from outside of the reported phase-space.
Analyzing one splitting per jet simplifies the correction chain, while retaining the calculability of the observable~\cite{vanBeekveld:2024wws}.  
The selected emissions from the PLJP at particle and detector levels are considered to be matched and fill the response matrix only if they are also the closest pair in the $y$--$\phi$ plane. 
This requirement suppresses off-diagonal elements in the response matrix caused by uncorrelated emissions, but it introduces the need for matching purity and efficiency corrections, which are applied before and after the unfolding, respectively.
In \PbPb collisions, the matching purity is of the order of 50\% at large angles and low \kt, where emissions created by the reclustering of the jet constituents with the ones from the UE dominate, and grows up to 80\% at high \kt. 
In \pp collisions, the purity is approximately 90\%. 
The matching efficiency is 90\% across the PLJP, except in the low-angle region where it falls to values of 60\%.

D'Agostini iterative unfolding with early stopping \cite{DAgostini:1994fjx} is used, as implemented in the \textsc{RooUnfold} 2.0.0 package \cite{RooUnfold}. 
The number of iterations is chosen such that the unfolded solution folded back to the detector level is statistically compatible with the input distribution by requiring a $p$-value~\cite{pvalue} of 0.95.
In \pp (\PbPb) collisions, the chosen iteration is 14 (23).
The unfolded angular distribution is normalized by the total number of jets, \Njets:
\begin{equation}
    \frac{1}{\Njets} \frac{\rd N_{\text{emissions}}}{\rd\lnD},
\end{equation}
and is reported in two \kt ranges: 10--20 and 20--40\GeV, determined by the unfolding uncertainty from below and by the available data statistics from above.

The measurement uncertainties are propagated in the unfolding procedure through variations of the response matrix and the purity and efficiency corrections, with the magnitude of the effect on observed distributions summarized in \appendixname~\ref{app:em}. 

To assess the physics model dependence of the regularized unfolding, different assumptions for the prior spectrum are used to cover a range of possibilities. 
In \pp collisions, the nominal MC sample generated by \PYTHIA is replaced by the one from \HERWIG. 
In \PbPb collisions, a modification of the particle-level jet emission pattern of the \PYTHIA events is considered.
To achieve this, the emission density of the whole PLJP (all primary emissions divided by the number of jets) is used as a proxy for the jet parton shower, and the \PYTHIA sample is reweighted in bins of \jetpt, emission \kt, and \lnD to match the emission pattern of two alternative models containing jet-quenching: \HYBRID~\cite{Casalderrey-Solana:2014bpa,Casalderrey-Solana:2020rsj} and \JEWEL 2.4.0~\cite{Zapp:2013vla}.
For both collision systems, the model uncertainty is determined by the symmetrized envelope formed by the variations compared to the nominal case.
To assess the impact of the individual energy calibration uncertainties of the PF candidates within jets, the four-momenta of the charged-particle, photon, and neutral-hadron PF candidates are independently scaled by $\pm$1, $\pm$3, and $\pm$5\% in simulation~\cite{CMS:2021iwu}, respectively. 
The uncertainty in the JES in \pp (\PbPb) collisions is around 3--4\% (4\%), increasing as a function of $\abs{\eta}$ (up to 10\% for \PbPb collisions in the barrel-endcap transition region $1.3 < \abs{\eta} < 1.6$).
An additional source of uncertainty in the JES is considered in \PbPb collisions to take into account the differences in the particle mixture in simulation and data caused by jet energy loss~\cite{CMS:2017ehl}.
In \pp collisions, the uncertainty in the resolution is in the range of 2--4\% in the barrel region, but is larger in the endcap and transition regions, where it varies in the range of 10--20\%.
To propagate the uncertainty from \jetpt resolution, the data-to-simulation resolution scale factors are varied according to their uncertainties, and the unfolding corrections are repeated.
The JES and JER uncertainties are also propagated to the hardest \kt subjet by scaling its energy by the same factor found for the full jet. 
In \PbPb collisions, there is an additional contribution to the JER uncertainty stemming from the UE modeling. The MC simulation is reweighted to match either the average background density, or the distribution of the number of tracks in the data. Both approaches lead to good descriptions of the fluctuations in the $\eta$--$\phi$ plane estimated in data and in simulation using the random cone method \cite{ALICE:2012nbx}. 
The uncertainty in the tracking efficiency is estimated by modifying 3 (4)\% of the tracks inside jets in \pp (\PbPb) collisions.
The four-momentum of the associated particle is smeared by 10\% to emulate the reconstruction of the particle using only calorimeter information, and in the case the smeared particle is within a distance $\sqrt{\smash[b]{(\Delta y)^2+(\Delta\phi)^2}}$ of 0.087 (the HCAL granularity) from a neutral hadron or a different charged particle, the four-momenta of the two particles are combined.
The statistical uncertainties in the MC sample used to construct the response matrix are considered as an additional systematic uncertainty.

\begin{figure*}
    \centering
    \includegraphics[width=0.49\linewidth]{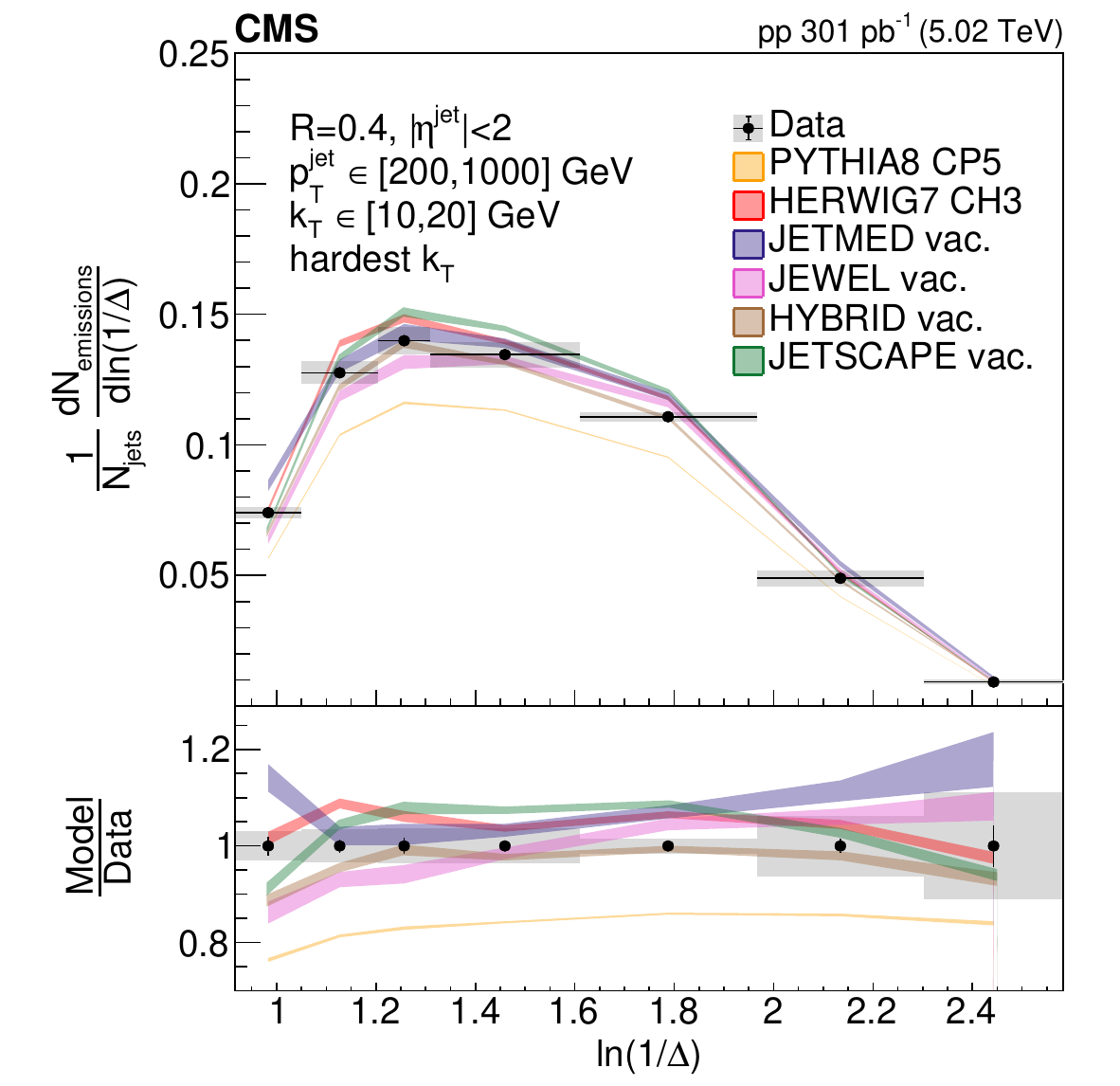}
    \includegraphics[width=0.49\linewidth]{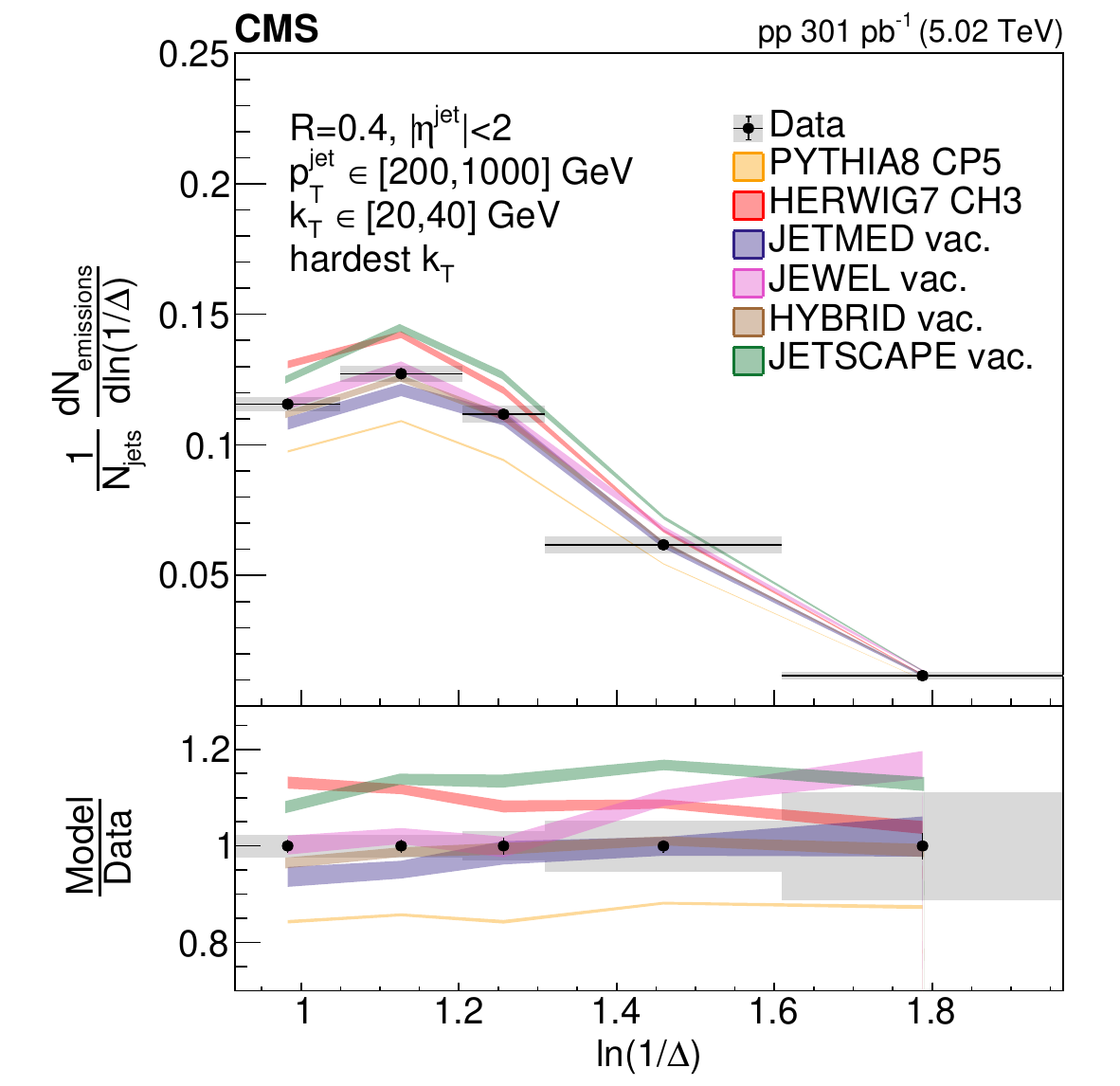}
    \caption{Particle-level angular distributions (\lnD) of the hardest \kt emissions in the ranges $\kt \in [10,20]\GeV$ (left) and $\kt \in [20,40]\GeV$ (right) in \pp collisions. A comparison is made with model predictions by \PYTHIA, \HERWIG, \JETMED, \JEWEL, \HYBRID and \JETSCAPE, with their ratios to data shown in the bottom panels. The uncertainties in the model predictions are statistical only. The shaded boxes represent the total uncertainty in data.}
    \label{fig:ppModelComparison}
\end{figure*}

The uncertainties are added in quadrature bin-by-bin for the final combination of systematic uncertainties. 
The uncertainties generally grow close to the diagonal edge of the PLJP, since this is the region where smearing effects are strongest and where the density of emissions falls steeply. Since the radiation pattern is modified in nontrivial ways in \PbPb collisions and since the \PbPb and \pp data were collected in different years, the systematic uncertainties are considered to be uncorrelated between the two measurements when computing their ratio.

The particle-level yield of the hardest primary emissions in \pp collisions in bins of \kt and \lnD is shown in Fig. \ref{fig:ppModelComparison}. 
The data are compared with \PYTHIA and \HERWIG simulations, as well as the vacuum radiation predictions of the jet quenching models \JETMED~\cite{Caucal:2020zcz}, \JEWEL, \HYBRID  and \JETSCAPE~\cite{JETSCAPE:2017eso}. 
The \JETMED sample is generated at the parton level with an angular-ordered shower. The \HYBRID vacuum baseline is the \PYTHIA generator using the Monash tune~\cite{Skands:2014pea}, with modifications to the final-state radiation \pt cutoff, and without multiple parton interactions. 
The \JETSCAPE predictions are using the \textsc{matter}~\cite{Matter1,Matter2} vacuum shower with the PP19 tune~\cite{PP19}.  
At high \kt, \JETMED and \HYBRID describe the data within~5\%, while \HERWIG and \JETSCAPE overestimate the number of emissions. 
At low \kt, all models apart from \PYTHIA show reasonable agreement with the data within ${\sim}15$\%.
The \PYTHIA predictions are consistently below the data in the number of emissions. This behavior was also observed in Ref.~\cite{CMS:2023lpp} and is attributed to the low final-state radiation value of the strong coupling constant $\alpS^{\text{FSR}}(\mZ^2)=0.118$ in the CP5 tune (where \mZ is the \PZ boson mass) compared to other tunes~\cite{CMS:2022kqg}.

\begin{figure*}
    \centering
    \includegraphics[width=0.49\textwidth]{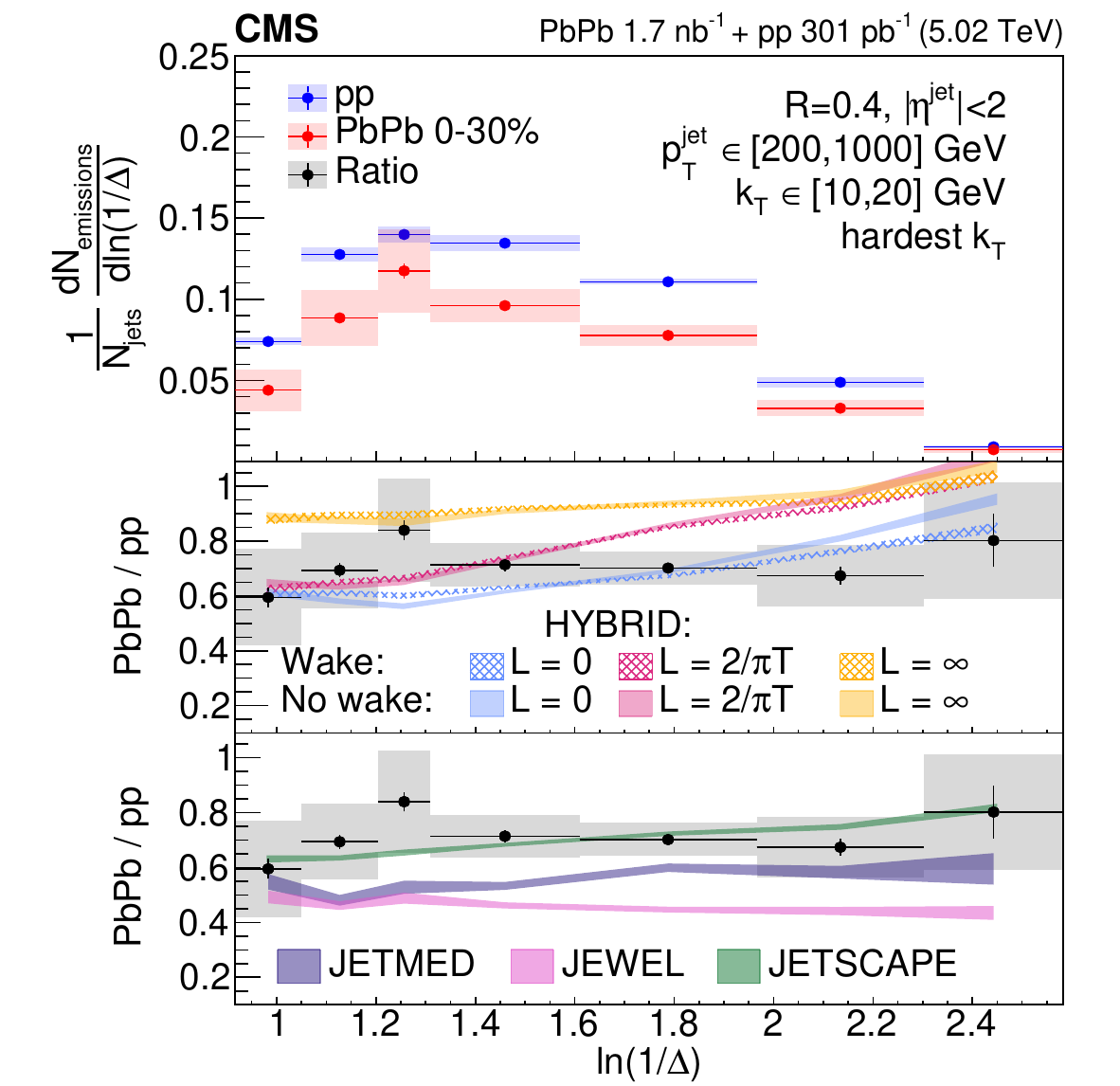}
    \includegraphics[width=0.49\textwidth]{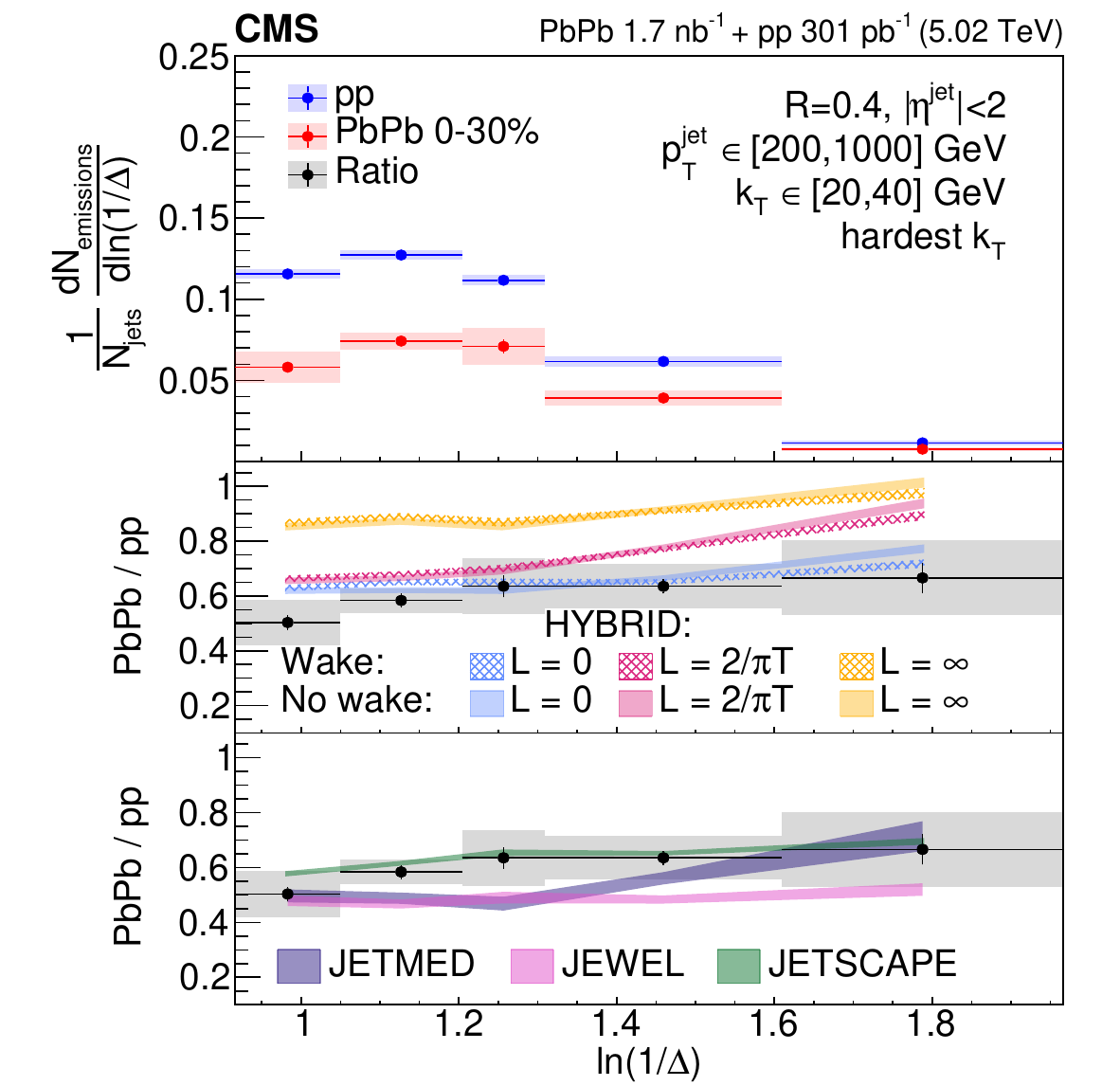}
    \caption{The hardest \kt emission angular distributions (\lnD) in the ranges $\kt \in [10,20]\GeV$ (left) and $\kt \in [20,40]\GeV$ (right) for \pp and \PbPb collisions of centrality 0--30\% and their ratios in the lower panels. The ratios are compared with jet quenching model predictions by \JETMED, \JETSCAPE, \JEWEL and \HYBRID, which provides different medium resolution lengths ($L$) and the inclusion/exclusion of backreacting medium particles (wake). The uncertainties in the model predictions are statistical only. The shaded boxes in the data distributions represent the total uncertainty.}
    \label{fig:RatioPlots}
\end{figure*}

The particle-level angular distribution of the hardest primary emission in \PbPb and \pp collisions, in bins of \lnD for different values of \kt, is shown in Fig.~\ref{fig:RatioPlots}. In the lower panels of the figure, the distribution ratio is compared with the one predicted by different physics models. 
A first observation is that the density of emissions is higher in \pp than in \PbPb collisions for $\kt>10\GeV$.
This is expected, since the jet radiation pattern is softened by the interaction with the medium, leading to lower overall jet and subjet \pt in \PbPb compared to \pp collisions.
Also, the distributions peak at larger angles (lower \lnD) in the $\kt\in[20,40]\GeV$ bin, as the phase space is more restricted at higher \kt due to its relation with \jetpt.
Finally, the shapes of the angular distributions in \PbPb and \pp are similar, resulting in ratios that are flat (\ie, independent of angle) within the experimental uncertainties. 
A $\chi^{2}$ test for fitting a flat line to the ratios in the higher \kt bin, utilizing the full covariance matrix, yields a $p$-value of 0.74.
The ratios for the lower \kt bin are also consistent with a flat line ($p$-value of 0.29), albeit with larger uncertainties.
The \pp jets fragment in vacuum, so finding identical shapes for the hardest intrajet emissions in \PbPb is compatible with the hypothesis that these earliest splittings are also produced in vacuum, after which the parton shower undergoes quenching with largely angle-independent energy loss.

Figure~\ref{fig:RatioPlots} also compares the ratio of the \PbPb and \pp distributions and four jet quenching models --- \HYBRID, \JETMED, \JETSCAPE, and \JEWEL. 
The first is a ``hybrid'' of weak and strong coupling description of medium-induced jet modifications. 
The model allows for the variation of the medium resolution length, $L$, above which charges of separation $(\Lmed-\tform)\Delta$ are resolved by the medium.
Three values of the parameter $L$ are considered: 0, \tpiT, and infinite ($\infty$), where $T$ is the temperature of the medium~\cite{Casalderrey-Solana:2014bpa}.
The $L=0$ case corresponds to the incoherent limit, in which the medium resolves all color charges individually.
The $L=\infty$ case corresponds to the fully coherent limit, in which the color charges within the jet are unresolved and the jet interacts with the medium as a single radiator. 
The value $L=\tpiT$ is a finite resolution length of the size of the Debye length in a weakly coupled plasma~\cite{Hulcher:2017cpt}.
The \JETMED model implements a perturbative QCD parton shower, including both factorized vacuum and medium-induced emissions.
The \JETSCAPE generator simulates jets in a multistage approach in medium using the model \textsc{matter+lbt} of \JETSCAPE v3.5 AA22 tune~\cite{JETSCAPE:2022jer,JETSCAPE:2022hcb,JETSCAPE:2023hqn}.
The \JEWEL generator treats the interplay of QCD radiation and rescattering in the QGP in a perturbative framework with fully microscopic dynamics.

The \HYBRID model with $L=0$ describes the data best in terms of the relative yield and shape of the ratio distribution.
The coherent limit ($L=\infty$) underestimates the suppression of emissions in \PbPb collisions, while the finite resolution case ($L=\tpiT$) shows steeper trends in the ratio than observed in the data. In all cases, the impact of the backreaction of the medium (or wake) is almost negligible, as expected in this high-\kt regime.
The $L=0$ curve is flatter in the higher-\kt bin.
The increase of steepness in the lower-\kt bin is potentially caused by the selection bias~\cite{CMS:2024zjn} caused by substructure-dependent energy loss.
The systematic uncertainties do not allow to conclude such a trend in the data.
The \JETMED model describes the ratio shape reasonably well, but slightly underestimates the absolute yield of emissions in the lower-\kt bin, possibly due to the lack of calibration of the energy loss in the model.
The \JEWEL model shows an almost perfectly flat ratio, indicating that the medium-induced hard scatterings implemented in the model do not contribute significantly to the \kt selection. The number of emissions is slightly underestimated, as in \JETMED.
The \JETSCAPE curves agree well with the data and show similar trends to those of the incoherent \HYBRID case. 

In summary, the primary Lund jet plane in heavy ion collisions has been presented for the first time in slices of relative momentum \kt.
This measurement provides an experimental approach to search for the presence of early vacuum-like emissions within jet evolution in heavy ion collisions.
The unfolded angular distribution of the hardest splittings selected in two intervals of \kt are compared in proton-proton (\pp) and lead-lead (\PbPb) collisions.
The ratio of the two distributions in data is compared with predictions by the \HYBRID, \JETSCAPE, \JEWEL, and \JETMED models.
Predictions employing both factorized and interleaved treatment of emissions are consistent with the observed ratio.
In the \kt ranges considered, the \HYBRID model distributions are unaffected by the inclusion of the backreaction of the medium.
The data ratios are most consistent with zero resolution length in the \HYBRID model and with the \JETSCAPE model, which does not have an explicit implementation of color coherence.
The \PbPb/\pp ratios in data are independent of angle within experimental uncertainties, with values smaller than unity.
This observation is consistent with the presence of early vacuum-like emissions in \PbPb collisions, which subsequently undergo quenching.

\begin{acknowledgments}
We thank Liliana Apolin{\'a}rio (\JEWEL), Paul Caucal (\JETMED), Daniel Pablos (\HYBRID), and Yasuki Tachibana (\JETSCAPE) for providing the corresponding model predictions in the kinematic range of the measurement and for discussions on their interpretation.

We congratulate our colleagues in the CERN accelerator departments for the excellent performance of the LHC and thank the technical and administrative staffs at CERN and at other CMS institutes for their contributions to the success of the CMS effort. In addition, we gratefully acknowledge the computing centers and personnel of the Worldwide LHC Computing Grid and other centers for delivering so effectively the computing infrastructure essential to our analyses. Finally, we acknowledge the enduring support for the construction and operation of the LHC, the CMS detector, and the supporting computing infrastructure provided by the following funding agencies: SC (Armenia), BMBWF and FWF (Austria); FNRS and FWO (Belgium); CNPq, CAPES, FAPERJ, FAPERGS, and FAPESP (Brazil); MES and BNSF (Bulgaria); CERN; CAS, MoST, and NSFC (China); MINCIENCIAS (Colombia); MSES and CSF (Croatia); RIF (Cyprus); SENESCYT (Ecuador); ERC PRG and PSG, TARISTU24-TK10 and MoER TK202 (Estonia); Academy of Finland, MEC, and HIP (Finland); CEA and CNRS/IN2P3 (France); SRNSF (Georgia); BMFTR, DFG, and HGF (Germany); GSRI (Greece); NKFIH (Hungary); DAE and DST (India); IPM (Iran); SFI (Ireland); INFN (Italy); MSIT and NRF (Republic of Korea); MES (Latvia); LMTLT (Lithuania); MOE and UM (Malaysia); BUAP, CINVESTAV, CONACYT, LNS, SEP, and UASLP-FAI (Mexico); MOS (Montenegro); MBIE (New Zealand); PAEC (Pakistan); MES, NSC, and NAWA (Poland); FCT (Portugal); MESTD (Serbia); MICIU/AEI and PCTI (Spain); MOSTR (Sri Lanka); Swiss Funding Agencies (Switzerland); MST (Taipei); MHESI (Thailand); TUBITAK and TENMAK (T\"{u}rkiye); NASU (Ukraine); STFC (United Kingdom); DOE and NSF (USA).
\end{acknowledgments}\section*{Data availability} Release and preservation of data used by the CMS Collaboration as the basis for publications is guided by the  \href{https://doi.org/10.7483/OPENDATA.CMS.1BNU.8V1W}{CMS data preservation, re-use and open access policy}.

\clearpage\bibliography{auto_generated}
\appendix
\section{Contributions to the measurement uncertainty}
\label{app:em}

Each considered source of uncertainty in the measurement and their contributions to the total fractional uncertainty are listed in Table~\ref{tab:Uncerts}.
\begin{table*}
    \renewcommand{\arraystretch}{1.4}
    \centering
    \topcaption{Lower and upper bounds of fractional uncertainties (in \%) for each source in \PbPb and \pp collisions for both reported \kt ranges.}
    \centering
    \begin{scotch}{lcccc}
     Uncertainty source & \multicolumn{2}{c}{$\kt \in [10,20]\GeV$} & \multicolumn{2}{c}{$\kt \in [20,40]\GeV$} \\
     \hline
    Collision system & \PbPb & \pp & \PbPb & \pp \\
    \hline
    Data (stat) & 2.7--12 & 0.61--3.0 & 4.0--8.1 & 0.85--2.0 \\
    MC (stat) & 1.3--3.5 & 0.54--1.4 & 1.8--2.9 & 0.69--1.1 \\
    JES & 0.95--9.5 & 0.094--0.87 & 1.1--5.1 & 0.81--1.6 \\
    JER & 0.37--4.2 & 0.14--0.85 & 1.0--5.0 & 0.14--0.90 \\
    Charged-particle energy scale & 2.4--5.7 & 0.48--6.4 & 1.7--6.0 & 0.51--6.7 \\
    Photon energy scale & 0.45--9.8 & 0.56--7.3 & 0.92--10 & 0.67--8.0 \\
    Neutral-hadron energy scale & 3.0--6.6 & 0.21--3.5 & 2.2--5.0 & 0.33--2.8 \\
    Tracking efficiency & 0.42--6.0 & $\leq$1.9 & 0.67--4.4 & $\leq$0.5 \\
    Physics model & 2.0--24 & 0.56--3.0 & 3.3--13 & 0.16--1.5 \\
    UE density & 0.67--7.3 & \NA & 0.48--4.3 & \NA \\
    Total & 8.5--29 & 1.6--11 & 7.3--17 & 2.4--11 \\
    \end{scotch}
    \label{tab:Uncerts}
\end{table*}\cleardoublepage \section{The CMS Collaboration \label{app:collab}}\begin{sloppypar}\hyphenpenalty=5000\widowpenalty=500\clubpenalty=5000\input{HIN-24-016-public-authorlist.tex}\end{sloppypar}
\end{document}

%% file: HIN-24-016-public-authorlist.tex
\cmsinstitute{Yerevan Physics Institute, Yerevan, Armenia}
{\tolerance=6000
A.~Belyaev\cmsorcid{0000-0003-1692-1173}, A.~Hayrapetyan, A.~Tumasyan\cmsAuthorMark{1}\cmsorcid{0009-0000-0684-6742}
\par}
\cmsinstitute{Institut f\"{u}r Hochenergiephysik, Vienna, Austria}
{\tolerance=6000
W.~Adam\cmsorcid{0000-0001-9099-4341}, J.W.~Andrejkovic, L.~Benato\cmsorcid{0000-0001-5135-7489}, T.~Bergauer\cmsorcid{0000-0002-5786-0293}, M.~Dragicevic\cmsorcid{0000-0003-1967-6783}, C.~Giordano\cmsorcid{0000-0001-6317-2481}, P.S.~Hussain\cmsorcid{0000-0002-4825-5278}, M.~Jeitler\cmsAuthorMark{2}\cmsorcid{0000-0002-5141-9560}, N.~Krammer\cmsorcid{0000-0002-0548-0985}, A.~Li\cmsorcid{0000-0002-4547-116X}, D.~Liko\cmsorcid{0000-0002-3380-473X}, M.~Matthewman, I.~Mikulec\cmsorcid{0000-0003-0385-2746}, J.~Schieck\cmsAuthorMark{2}\cmsorcid{0000-0002-1058-8093}, R.~Sch\"{o}fbeck\cmsAuthorMark{2}\cmsorcid{0000-0002-2332-8784}, D.~Schwarz\cmsorcid{0000-0002-3821-7331}, M.~Shooshtari\cmsorcid{0009-0004-8882-4887}, M.~Sonawane\cmsorcid{0000-0003-0510-7010}, W.~Waltenberger\cmsorcid{0000-0002-6215-7228}, C.-E.~Wulz\cmsAuthorMark{2}\cmsorcid{0000-0001-9226-5812}
% ...
\par}
\vskip\cmsinstskip
\dag:~Deceased\\
$^{1}$Also at Yerevan State University, Yerevan, Armenia\\
$^{2}$Also at TU Wien, Vienna, Austria\\
% ...
